\documentclass[11pt,prd,a4paper,preprintnumbers,amsmath,amssymb,nofootinbib]{article}
\pdfoutput=1
\usepackage{feynmp-auto,expdlist}
\usepackage{amsmath,amsfonts,amssymb}
\usepackage{graphicx}
\usepackage{enumerate}
\usepackage{hyperref}
\usepackage{latexsym}
\usepackage{hepnicenames}
\usepackage{enumerate}
\usepackage{soul}
\usepackage[normalem]{ulem}
\usepackage{wasysym}
\usepackage{makecell}
\usepackage{bbm}
\usepackage{physics}
\usepackage[version=4]{mhchem}
\usepackage{amsmath}

\usepackage{comment}
\usepackage{slashed}

\font\tenrsfs=rsfs10 at 12pt
\font\sevenrsfs=rsfs7
\font\fiversfs=rsfs5
\newfam\rsfsfam
\textfont\rsfsfam=\tenrsfs
\scriptfont\rsfsfam=\sevenrsfs
\scriptscriptfont\rsfsfam=\fiversfs

\numberwithin{equation}{section}

\usepackage{mathrsfs}
\usepackage{braket}
\usepackage{titling}
\usepackage{amsmath}
\usepackage{slashed}
\usepackage{amssymb}
\usepackage{epsfig}
\usepackage{graphicx}
\usepackage{color}
\usepackage{rotating}
\usepackage[margin=1.in]{geometry}
\usepackage[table,xcdraw,dvipsnames]{xcolor}
\usepackage[compress,numbers,sort]{natbib}
\usepackage{colortbl}
\usepackage{pdflscape}
\usepackage{color}
\usepackage{mathtools}
\usepackage{comment}
\usepackage{multirow}
\usepackage{booktabs}

\definecolor{nicered}{rgb}{0.7,0.1,0.1}
\definecolor{nicegreen}{rgb}{0.1,0.5,0.1}
\definecolor{red}{rgb}{1.0, 0, 0}
\definecolor{niceblue}{rgb}{0,0,0.8}
\allowdisplaybreaks

\definecolor{blus}{cmyk}{1,1,0,0.6}
\definecolor{verde}{cmyk}{0.92,0,0.59,0.25}
\definecolor{rossos}{cmyk}{0,1,1,0.55}

\hypersetup{colorlinks,bookmarksopen,bookmarksnumbered,linkcolor=blus,pdfstartview=FitH,urlcolor=rossos,citecolor=verde}

\newcommand{\be}{\begin{equation}}
\newcommand{\ee}{\end{equation}}
\newcommand{\bea}{\begin{eqnarray}}
\newcommand{\eea}{\end{eqnarray}}
\newcommand{\bei}{\begin{itemize}}
\newcommand{\eei}{\end{itemize}}

\newcommand{\nn}{\nonumber}

\newcommand{\eff}{{\rm eff}}

\newcommand{\mcL}{\mathcal L}
\def\L{\mathscr L}

\newcommand{\beq}{\begin{equation}}
\newcommand{\eeq}{\end{equation}}

\renewcommand{\[}{\left[}

\renewcommand{\(}{\left(}
\renewcommand{\)}{\right)}

\newcommand{\bkg}{{\rm bkg}}
\newcommand{\rec}{{\rm rec}}
\newcommand{\bin}{{\rm bin}}
\newcommand{\obs}{{\rm obs}}

\renewcommand{\th}{{\rm th}}

\def\inv{{\rm inv}}

\newcommand{\SU}{{\rm SU}}

\newcommand{\U}{{\rm U}}

\newcommand{\1}{{\textbf{1}}}

\newcommand{\3}{{\textbf{3}}}

\newcommand{\eps}{\epsilon}

\newcommand{\wt}[1]{\widetilde{#1}}

\newcommand{\lrpartial}[1]{\overset{\leftrightarrow}{\partial_{#1}}}

\newcommand{\BR}{{\cal B }}

\def\vev#1{\langle #1\rangle}

\begin{document}

\begin{flushright}
KYUSHU-HET-349 \\
LAPTH-004/26
\end{flushright}

\vspace{0.8cm}

\begin{center}  
{\LARGE
\bf\color{blus}
Dark Matter emission at Belle~II and NA62\\
\vspace{0.1cm}
in Minimal Flavor Violation framework
}
\vspace{0.8cm}

{\bf 
Federico Mescia$^{a}$, 
Shohei Okawa$^{b,c,d}$, 
Joel Swallow$^{e}$, 
Claudio Toni$^{f}$ }\\[5mm]

{\it $^a$Istituto Nazionale di Fisica Nucleare, Laboratori Nazionali di Frascati,
C.P. 13, 00044 Frascati, Italy}\\[1mm]
{\it $^b$Asia Pacific Center for Theoretical Physics, Pohang, 37673, Korea}\\[1mm]
{\it $^c$Department of Physics, Pohang University of Science and Technology, Pohang, 37673, Korea}\\[1mm]
{\it $^d$Department of Physics, Kyushu University, 744 Motooka, Nishi-ku, Fukuoka, 819-0395, Japan}\\[1mm]
{\it $^e$CERN, European Organization for Nuclear Research, CH-1211 Geneva 23,
Switzerland}\\[1mm]
{\it $^f$LAPTh, Université Savoie Mont-Blanc et CNRS,
74941 Annecy, France}

\vspace{0.3cm}
\begin{quote}

Minimal Flavor Violation (MFV) provides a compelling framework for  exploring physics beyond the Standard Model, in which new QCD-singlet fields transforming under the global $\mathrm{SU}(3)^3$ quark flavor symmetry can naturally be stable and act as dark matter (DM) candidates.
We show that the DM--MFV framework naturally accommodates the excess in either $K^+ \to \pi^+ \nu \bar{\nu}$ or $B^+ \to K^+ \nu \bar{\nu}$, while a unified explanation of both channels simultaneously cannot be achieved within a minimal setup containing only a single dark matter multiplet with nearly degenerate masses.
Overall, our findings underscore the intricate interplay between MFV-based model building, flavored dark matter scenarios, and precision flavor experiments, highlighting flavored dark matter as a framework that is both theoretically robust and experimentally testable.

\end{quote}

\thispagestyle{empty}

\end{center}
\newpage
\setcounter{tocdepth}{2}
\tableofcontents

\section{Introduction}

Rare flavor-changing neutral current (FCNC) processes such as those involving $s \to d$ and $b \to s$ transitions are among the most sensitive probes of physics beyond the Standard Model (BSM). Their theoretical cleanliness---stemming from the dominance of short-distance dynamics, and the absence of large hadronic uncertainties---makes them ideal for testing new physics scenarios. 

Experimental efforts by NA62~\cite{NA62:2024pjp} and KOTO~\cite{KOTO:2024zbl} in Kaon decays, and Belle~II~\cite{Belle-II:2023esi}
in $B$-meson decays, have significantly improved bounds on $K \to \pi \nu \bar{\nu}$ and $B \to K^{(*)} \nu \bar{\nu}$ channels, respectively. While the Standard Model (SM) predicts a coherent pattern for these processes, emerging data provide mild yet intriguing hints of deviations. In particular, the central value of the $K^+ \to \pi^+ \nu \bar{\nu}$ branching ratio reported by NA62 lies somewhat above the SM prediction~\cite{NA62:2024pjp}. Even more strikingly, the latest Belle~II results in the $B^+ \to K^+ \nu \bar{\nu}$ channel show an excess with a significance of about $2.7\sigma$~\cite{Belle-II:2023esi}.

Many studies have investigated new physics interpretations of the recent Belle II excess in $B^+\to K^+ \nu \bar\nu$ \cite{Belle-II:2023esi}, mostly in two approaches. 
In the first approach, the excess is attributed to heavy new physics. 
Integrating out new heavy degrees of freedom induces various effective $bs\nu\nu$ interaction  operators, which provide additional contributions to the $b \to s \nu\bar\nu$ transition.
Effective field theories (EFTs), with or without light right handed neutrinos, are employed to perform a model-independent analysis in this scenario~\cite{Athron:2023hmz, Bause:2023mfe, Allwicher:2023xba, Felkl:2023ayn, He:2023bnk, Chen:2024jlj, Hou:2024vyw, Marzocca:2024hua, Rosauro-Alcaraz:2024mvx, Buras:2024ewl, Kim:2024tsm, Allwicher:2024ncl, Becirevic:2024iyi, Lee:2025jky}, allowing one to identify distinct correlations among various flavor observables. 
In particular, correlations among $\BR(B^+ \to K^+ \nu\bar\nu)$, $\BR(B\to K^* \nu \bar\nu)$, and longitudinal polarization fraction of $K^*$ in $B\to K^* \nu \bar\nu$ helps disentangle different new physics contributions to $B\to K^{(*)}\nu\bar\nu$ observables~\cite{Bause:2023mfe, Allwicher:2023xba, Hou:2024vyw, Rosauro-Alcaraz:2024mvx, Buras:2024ewl}. 
Concrete ultraviolet realizations are also proposed, including a supersymmetric model~\cite{Wang:2023trd}, models with leptoquarks~\cite{Chen:2023wpb, Marzocca:2024hua, Hati:2024ppg, Zhang:2024rmb, Chen:2025npb, Crivellin:2025qsq, Becirevic:2024iyi} and $Z'$ bosons~\cite{Athron:2023hmz, Marzocca:2024hua, Buras:2024mnq}, scotogenic model~\cite{Chen:2024cll}, grand unified theory \cite{Bhattacharya:2024clv}, and universal extra-dimension model \cite{Shaw:2025ays}.

In the second approach, the Belle~II excess is interpreted as dark particle emission in $B \to K$ decays: two-body ($B \to K X$) or three-body ($B \to K \phi\phi$) processes. 
The emitted dark particle can be a scalar, fermion, or vector, and may act as dark matter.
The studied dark particles include 
a singlet (Higgs-portal) scalar~\cite{Abdughani:2023dlr, Berezhnoy:2023rxx, Datta:2023iln, McKeen:2023uzo, Fridell:2023ssf, Ho:2024cwk, He:2024iju, Bolton:2024egx, He:2025jfc, Berezhnoy:2025tiw, Bolton:2025fsq, Ko:2025drr, Berezhnoy:2025nmb, Kim:2025zaf}, 
axion and axion-like particles~\cite{Altmannshofer:2023hkn, Altmannshofer:2024kxb, Hu:2024mgf, Calibbi:2025rpx, Ding:2025eqq, Li:2025ski, Gao:2025ohi, Alda:2025uwo, Abumusabh:2025zsr}, 
a singlet fermion~\cite{Fridell:2023ssf, Bolton:2024egx, Bolton:2025fsq},
dark photon~\cite{Gabrielli:2024wys, Calibbi:2025rpx}, 
and light vector bosons~\cite{Altmannshofer:2023hkn, Fridell:2023ssf, Bolton:2024egx, Hu:2024mgf, Aliev:2025hyp, Bolton:2025fsq, DiLuzio:2025qkc, Bolton:2025lnb, Berezhnoy:2025nmb}. 
According to a model-agnostic analysis for the missing invariant mass squared ($q^2$) distribution~\cite{Fridell:2023ssf, Bolton:2024egx, Bolton:2025fsq}, 
two-body and three-body scenarios both provide a good description for the excess, 
with the best fit for $M_X \simeq 2$\,GeV and $M_\phi\simeq0.5$\,GeV for the two-body and three-body scenarios, respectively.

In this paper, we examine whether this Belle~II $B^{+} \to K^{+}\nu\bar{\nu}$ excess, together with the slightly higher-than-expected $K^{+}\rightarrow\pi^{+}\nu\bar{\nu}$ branching ratio measured by NA62~\cite{NA62:2024pjp}, can be accommodated within the \emph{Minimal Flavor Violation} (MFV) dark matter framework~\cite{Batell:2011tc,Lopez-Honorez:2013wla,Batell:2013zwa,Bishara:2015mha,Mescia:2024rki}. 
For $B^{+} \to K^{+}\nu\bar{\nu}$, our analysis is similar to the one outlined in~\cite{Bolton:2024egx,Bolton:2025lnb,Bolton:2025fsq}, and we expand this to search for a simultaneous description of this excess, together with available data on other $d_{i}\rightarrow d_{j}\nu\bar{\nu}$ processes.

MFV~\cite{Chivukula:1987py, Hall:1990ac, Buras:2000dm, DAmbrosio:2002vsn} postulates that all sources of flavor and CP violation originate from the Standard Model Yukawa couplings, ensuring that any new physics sector obeys the same flavor structure as the SM. In this setup, new operators contributing to $s \to d \nu \bar{\nu}$ and $b \to s \nu \bar{\nu}$ transitions are aligned with Cabibbo–Kobayashi–Maskawa (CKM) hierarchies, naturally generating correlated predictions for Kaon and $B$-meson observables.
Strikingly, MFV symmetries can stabilize additional flavor multiplets, allowing them to play the role of dark matter.
It has previously been shown that in MFV the lightest state of a new QCD-singlet field $\chi$ that transforms under the quark flavor subgroup $\SU(3)_q \times \SU(3)_u \times \SU(3)_d$ is stable, even when all higher dimensional operators are included~\cite{Mescia:2024rki,Batell:2011tc}. 
This is due only to the invariance under the color and flavor groups within the MFV.
Remarkably, the addition of the $\chi$ fields, acting as spurions of the flavor group beyond the usual SM Yukawa matrices, breaks the minimally expected scaling of BSM contribution in strict U(3)$^3$ MFV: $C^\text{BSM}_{ij} \sim (Y_u^\dagger Y_u)_{ij}\propto V^*_{ti}V_{tj}$.
Investigation of MFV beyond minimal assumptions is therefore well motivated, in particular considering implications for $d_{i}\rightarrow d_{j}\chi_i\bar{\chi}_j$ processes, where $\chi_i$ has the same flavor charge as the quark fields.

Here, we show that a unified explanation of both anomalies cannot be achieved within a minimal setup containing only a single DM multiplet. Accommodating both channels simultaneously requires introducing an additional DM multiplet with a different mass for each flavor. 
In our analysis we construct a likelihood which contains all the publicly available experimental data, searching for a comprehensive phenomenological description of this scenario.

This paper is organized as follows. In Sec.~\ref{sec:MFVDM}, we review effective operators for MFV DM. Sec.~\ref{sec:exp} presents the experimental setup for rare $K$ and $B$ decays. Our main results are given in Sec.~\ref{sec:results} and we conclude in Sec.~\ref{sec:conclusion}. Technical details and calculation tools are provided in the appendices.

\section{Dark Matter within Minimal Flavor Violation}
\label{sec:MFVDM}

The SM exhibits a large global flavor symmetry $\U(3)^5=\U(3)_q\times \U(3)_u\times \U(3)_d\times\U(3)_\ell\times \U(3)_e$ in the gauge sector~\cite{Chivukula:1987py, Gerard:1982mm}. 
This symmetry is explicitly broken by the Yukawa interactions of quarks and leptons to the doublet Higgs field $H$, and flavor violating processes show specific patterns controlled by the structure of the Yukawa interaction matrices $Y_{u,d,e}$. 

New interactions from physics beyond the SM can introduce independent sources of flavor violation, generally at very high scales much above the TeV scale. The MFV hypothesis~\cite{Chivukula:1987py, Hall:1990ac, Buras:2000dm, DAmbrosio:2002vsn} requires that new physics interactions also respect the $\U(3)^5$ flavor symmetry, with breaking controlled solely by the Yukawa matrices.
Formally, MFV is implemented by promoting the Yukawa matrices to spurion fields which transform like 
\beq
Y_u \sim ({3},\bar{{3}},{1},{1},{1}) \,,\quad
Y_d \sim ({3},{1},\bar{{3}},{1},{1}) \,,\quad
Y_e \sim ({1},{1},{1},{3},\bar{{3}}) \,,
\eeq
under the $\U(3)^5$ flavor group.

In this formalism, for any operator ${\cal O}_{ij\dots}$ with flavor indices $i,j,\dots$, its coupling $C_{ij\dots}$ is expanded in Yukawa spurions to ensure invariance under flavor transformations. 
More concretely, $C_{ij}$ is expressed for ${\cal O}_{ij}\sim({1},{3}\otimes\bar{3},{1},{1},{1})$ as 
\beq
C_{ij} = c_0\,\delta_{ij} + \epsilon \, c_1 (Y_u^\dagger Y_u)_{ij} 
    + \epsilon^2 \Big[c_2 (Y_u^\dag Y_u Y_u^\dag Y_u)_{ij} 
    + c_2^\prime (Y_u^\dag Y_d Y_d^\dag Y_u)_{ij}\Big] + \dots \,,
\eeq
where the ellipsis represents further spurion insertions, which are suppressed by additional powers of Yukawa couplings, CKM off-diagonal elements, and a potentially small MFV expansion parameter $\epsilon$.

Remarkably, when new QCD-singlet particles are added to the SM, the principle of MFV can also ensure the stability of some of them, allowing these particles to serve as dark matter (DM) candidates. 
Consider a new QCD-singlet field $\chi$, transforming like $\chi\sim(n_q, m_q)\times(n_u, m_u)\times(n_d, m_d)$ under $G_q=\SU(3)_q \times \SU(3)_u \times \SU(3)_d$. 
One can show that the lightest state of $\chi$ is absolutely stable when $(n_q+n_u+n_d-m_q-m_u-m_d)\,{\rm mod}\,3\neq0$ is satisfied~\cite{Batell:2011tc, Batell:2013zwa}.
This state, $\chi$, is thus an excellent DM candidate if electrically neutral. Table~\ref{tab:DMcandidates} lists the lowest-dimensional representations of $G_q$ that are stable according to the above conditions.

\begin{table}[h!]
\centering
\renewcommand{\arraystretch}{1.4}
\begin{tabular}{|c|c|}
\hline
$\,(n,m)\,$ & $SU(3)_q \times SU(3)_u \times SU(3)_d$ \\ \hline\hline


$(1,0)$ & $({3}, {1}, {1}),\; ({1}, {3}, {1}),\; ({1}, {1}, {3})$ \\ \hline

$(0,1)$ & $(\overline{{3}}, {1}, {1}),\; ({1}, \overline{{3}}, {1}),\; ({1}, {1}, \overline{{3}})$ \\ \hline

$(2,0)$ &
$({6}, {1}, {1}),\; ({1}, {6}, {1}),\; ({1}, {1}, {6})$\\
& $({3}, {3}, {1}),\; ({3}, {1}, {3}),\; ({1}, {3}, {3})$ \\ \hline

$(0,2)$ &
$(\overline{{6}}, {1}, {1}),\; ({1}, \overline{{6}}, {1}),\; ({1}, {1}, \overline{{6}})$\\
& $(\overline{{3}}, \overline{{3}}, {1}),\; (\overline{{3}}, {1}, \overline{{3}}),\; ({1}, \overline{{3}}, \overline{{3}})$ 
\\ \hline
\end{tabular}
\caption{ Flavored DM candidates. Listed are the lowest-dimensional 
  representations of $G_q$ that are stable 
  once MFV is imposed, with $(n,m)$ where $n \equiv n_q+n_u+n_d$, 
  $m\equiv m_q + m_u + m_d$.  Depending on their electroweak quantum numbers, these multiplets 
  may contain viable DM candidates.
}
\label{tab:DMcandidates}
\end{table}
In this framework, interactions between quarks and dark matter particles determine the phenomenology in the dark sector: e.g. dark matter production in the early universe, for (in)direct searches, and decays of the heavy dark states. 
The traditional Weakly Interacting Massive Particle (WIMP) parameter space, where $\chi$ is heavier than $B$ mesons, was comprehensively studied in~\cite{Batell:2011tc, Lopez-Honorez:2013wla}, with implications ranging from freeze-out cosmology to collider signals and FCNC processes. 
The possibility of multi-component dark matter was investigated in \cite{Mescia:2024rki}.
If instead $\chi$ is lighter than $B$ mesons, some operators can induce dark matter emission from heavy quark decays, such as $d_i \to d_j \chi_i \bar{\chi}_j$. 
Since $\chi$ must be effectively unobservable with current detectors, these new decay modes mimic the FCNC decays, $d_i \to d_j \nu\bar{\nu}$. 

In this section, we classify the leading effective operators contributing to $d_i \to d_j \chi_i \bar{\chi}_j$, with a focus on the lowest-dimensional representations with $(n,m)=(1,0)$. 
In this work, we consider a flavored dark matter multiplet $\chi=S$ (scalar) or $\psi$ (fermion) as the only new light degrees of freedom below the electroweak scale.
Their interactions with the SM quarks are described by higher‑dimensional operators whose flavor structure is fixed by the MFV hypothesis. In this framework, the fields $\chi$ are degenerate at leading order in the MFV expansion, and mass splittings arise only from insertions of the Yukawa spurions at higher order:
\beq
(m_S^2)_{ij} = m_S^2 \left[ \delta_{ij} + c_S \, \epsilon \, (Y_q^\dagger Y_q)_{ij} + \dots \right]
\eeq
for a scalar multiplet $S$, or 
\beq
(m_\psi)_{ij} = m_\psi \left[ \delta_{ij} + c_\psi \, \epsilon \, (Y_q^\dagger Y_q)_{ij} + \dots \right]
\eeq
for a fermion multiplet $\psi$.
Here, $c_{S,\psi}$ are unknown coefficients of order unity, and $\epsilon \leq 1$ counts the number of Yukawa insertions in the MFV expansion. The matrix $Y_q$ denotes either $Y_u$ or $Y_d$, depending on the flavor representation of the new fields.
Throughout this paper, we ignore small corrections of ${\cal O}(\epsilon)$ and only keep the leading order terms. Consequently, the $\chi$ fields are degenerate at this order.

\subsection{Scalar DM}

First, we focus on the case of scalar DM $S$. The relevant interaction Lagrangian is composed of dimension-6 operators involving two quark fields and two dark matter fields, 
\beq
\L^\eff_S = \frac{1}{\Lambda^2} \sum_I \sum_{i,j,k,l} C_{ijkl}^I {\cal O}_{ijkl}^I \,,
\eeq
where $i,j,k,l=1,2,3$ are quark flavor indices and a full set of the operators for scalar DM is given by 
\begin{align}
{\cal O}^1_{ijkl} & = (\bar{q}_{Li} \gamma^\mu q_{Lj}) (S^\dagger_k i \lrpartial{\mu} S_l) \,,\\
{\cal O}^2_{ijkl} &= (\bar{u}_{Ri} \gamma^\mu u_{Rj}) (S^\dagger_k i \lrpartial{\mu} S_l) \,,\\
{\cal O}^3_{ijkl} & = (\bar{d}_{Ri} \gamma^\mu d_{Rj}) (S^\dagger_k i \lrpartial{\mu} S_l) \,,\\
{\cal O}^4_{ijkl}  = (\bar{q}_{Li} \wt{H} u_{Rj}) (S^\dagger_k S_l) & \,,\quad
{\cal O}^{\overline{4}}_{ijkl} = (\bar{u}_{Ri} \wt{H}^\dagger q_{Lj}) (S^\dagger_k S_l) = \left[{\cal O}^4_{jilk}\right]^\dagger \,,\\
{\cal O}^5_{ijkl} = (\bar{q}_{Li} H d_{Rj})(S^\dagger_k S_l) & \,,\quad
{\cal O}^{\overline{5}}_{ijkl} = (\bar{d}_{Ri} H^\dagger q_{Lj}) (S^\dagger_k S_l) = \left[{\cal O}^5_{jilk}\right]^\dagger \,,
\end{align}
where $A\lrpartial{\mu} B = A (\partial_\mu B) - (\partial_\mu A) B$.
Hermiticity requires that
\be
\begin{split}
C_{ijkl}^1  =\left[C_{jilk}^1\right]^*  \,,\quad
C_{ijkl}^2  &=\left[C_{jilk}^2\right]^* \,,\quad
C_{ijkl}^3  =\left[C_{jilk}^3\right]^*  \,,\\
C_{ijkl}^4 =\left[C_{jilk}^{\overline{4}}\right]^* & \,,\quad
C_{ijkl}^5  =\left[C_{jilk}^{\overline{5}}\right]^* \,.
\end{split}
\ee
Of these operators, ${\cal O}^1_{ijkl}$, ${\cal O}^3_{ijkl}$ and ${\cal O}^5_{ijkl}$ provide flavor violating interactions to the down quarks at the tree level, so that hereafter we only consider these three operators. 

Under MFV, the coefficients $C_{ijkl}^I$ are expanded in terms of the quark Yukawa matrices and their concrete expressions depend on the $G_q$ representation of the DM field $S_i$. 

\subsubsection{$S\sim(\3,\1,\1)$}

The leading coefficients are given by 
\begin{align}
\label{eq:cS311_1}
C_{ijkl}^1 & = c^1_1 \delta_{ij} \delta_{kl} + c^1_2 \delta_{il} \delta_{jk} + \dots \ , \\
\label{eq:cS311_3}
C_{ijkl}^3 & = c^3_1 \delta_{ij} \delta_{kl} + \dots \ , \\
\label{eq:cS311_5}
C_{ijkl}^5 & = c^5_1 (Y_d)_{ij} \delta_{kl} + c^5_2 (Y_d)_{kj} \delta_{il} + \dots \ ,
\end{align}
where the ellipses indicate further Yukawa matrix insertions and $c^n_m$ are taken to be real.

Without loss of generality, one can take $(Y_d)_{ij}=y_d^i \delta_{ij}$ and $Y_u=V^\dagger \hat{Y}_u$, with $V$ being the CKM matrix and $(\hat{Y}_u)_{ij} = y_u^i \delta_{ij}$.
In the unitary gauge $\vev{H}=(0,v/\sqrt{2})^T$ and 
in the quark mass basis ($u_L \to V^\dagger u_L$), we find 
\begin{align}
\L^\eff_S \supset 
\frac{1}{\Lambda^2} \sum_{i,j}\Bigg\{& 
    c^1_2 (\bar{d}_{Li} \gamma^\mu d_{Lj})(S^\dagger_j i \lrpartial{\mu} S_i)
   + c^1_2 \sum_{k,l} V_{il} V^*_{jk}( \bar{u}_{Li} \gamma^\mu u_{Lj}) (S^\dagger_k i \lrpartial{\mu} S_l) \nonumber \\
    \label{eq:LS311}
    +& c^5_2 (m_d^j+m_d^i) (\bar{d}_{i} d_{j}) (S^\dagger_j S_i)+ c^5_2 (m_d^j-m_d^i) (\bar{d}_{i} \gamma_5 d_{j}) (S^\dagger_j S_i) \Bigg\} \, ,
\end{align}
where $m_d^i = y_d^i v /\sqrt{2}$.
Note that in the last equation both quarks and DM fields are in the mass basis.

\subsubsection{$S\sim(\1,\3,\1)$}

The leading coefficients are given by 
\begin{align}
\label{eq:cS131_1}
C_{ijkl}^1 & = c^1_1 \delta_{ij} \delta_{kl} + \dots \ , \\
\label{eq:cS131_3}
C_{ijkl}^3 & = c^3_1 \delta_{ij} \delta_{kl} + \dots \ , \\
\label{eq:cS131_5}
C_{ijkl}^5 & = c^5_1 (Y_d)_{ij} + \dots \ ,
\end{align}
where the ellipses indicate further Yukawa matrix insertions. 
There is no flavor violating interaction in the down sector, thus $S\sim(1,3,1)$ is not a suitable representation to address the Belle~II and NA62 excesses.

\subsubsection{$S\sim(\1,\1,\3)$}

The leading coefficients are given by 
\begin{align}
\label{eq:cS113_1}
C_{ijkl}^1 & = c^1_1 \delta_{ij} \delta_{kl} + \dots \ , \\
\label{eq:cS113_3}
C_{ijkl}^3 & = c^3_1 \delta_{ij} \delta_{kl} + c^3_2 \delta_{il} \delta_{jk} + \dots \ ,\\
\label{eq:cS113_5}
C_{ijkl}^5 & = c^5_1 (Y_d)_{ij} \delta_{kl} + c^5_2 (Y_d)_{il} \delta_{jk} + \dots \ ,
\end{align}
where the ellipses indicate further Yukawa matrix insertions and $c^n_m$ are taken to be real. 
The flavor violating interactions arise only from ${\cal O}^3_{ijkl}$ and ${\cal O}^5_{ijkl}$, which are in the form 
\begin{align}
\L^\eff_S \simeq \frac{1}{\Lambda^2} \sum_{i,j}\Bigg\{& 
    c^3_2 \(\bar{d}_{Li} \gamma^\mu d_{Lj} \) (S^\dagger_j i \lrpartial{\mu} S_i)
    \nn \\
    +& c^5_2 (m_d^i+m_d^j) (\bar{d}_{i} d_{j}) (S^\dagger_j S_i)+ c^5_2 (m_d^i-m_d^j) (\bar{d}_{i} \gamma_5 d_{j}) (S^\dagger_j S_i) \Bigg\} \, ,
\end{align}
where we take the unitary gauge and the quark mass basis. Note that the $d_i\to d_j$ operators of this Lagrangian are identical to those of Eq.~\eqref{eq:LS311} with $c^1_2\to c^3_2$.

\subsection{Fermion DM}

Next, we focus on the case of fermion DM $\psi$.
The effective Lagrangian consists of a set of higher dimensional operators, 
\beq
\L^{\eff}_F = \sum_I \frac{1}{\Lambda^{d-4}} \, C_{ijkl}^I {\cal O}_{ijkl}^I \,,
\eeq
where $d$ denotes the dimension of the operator ${\cal O}^I_{ijkl}$. 
The dimension-6 operators for fermion DM are given by 
\begin{align}
{\cal O}^{1V}_{ijkl} & = (\bar{q}_{Li} \gamma^\mu q_{Lj}) (\bar{\psi}_k \gamma_\mu \psi_l) \,,\quad
{\cal O}^{1A}_{ijkl} = (\bar{q}_{Li} \gamma^\mu q_{Lj}) (\bar{\psi}_k \gamma_\mu \gamma_5 \psi_l) \,,\\
{\cal O}^{2V}_{ijkl} & = (\bar{u}_{Ri} \gamma^\mu u_{Rj}) (\bar{\psi}_k \gamma_\mu \psi_l) \,,\quad
{\cal O}^{2A}_{ijkl} = (\bar{u}_{Ri} \gamma^\mu u_{Rj}) (\bar{\psi}_k \gamma_\mu \gamma_5 \psi_l) \,,\\
{\cal O}^{3V}_{ijkl} & = (\bar{d}_{Ri} \gamma^\mu d_{Rj}) (\bar{\psi}_k \gamma_\mu \psi_l) \,,\quad
{\cal O}^{3A}_{ijkl} = (\bar{d}_{Ri} \gamma^\mu d_{Rj}) (\bar{\psi}_k \gamma_\mu \gamma_5 \psi_l) \,.
\end{align}
The scalar-scalar operators are now dimension-7, and are given by
\begin{align}
{\cal O}^{4S}_{ijkl}  = (\bar{q}_{Li} \wt{H} u_{Rj}) (\bar{\psi}_k \psi_l) \,,&\quad
{\cal O}^{\overline{4S}}_{ijkl} = (\bar{u}_{Ri} \wt{H}^\dagger q_{Lj}) (\bar{\psi}_k \psi_l) = \left[{\cal O}^{4S}_{jilk}\right]^\dagger \,,\\
{\cal O}^{4P}_{ijkl}  = (\bar{q}_{Li} \wt{H} u_{Rj})
(\bar{\psi}_k i\gamma_5 \psi_l) \,,&\quad
{\cal O}^{\overline{4P}}_{ijkl} = (\bar{u}_{Ri} \wt{H}^\dagger q_{Lj})(\bar{\psi}_k i\gamma_5 \psi_l) = \left[{\cal O}^{4P}_{jilk}\right]^\dagger \,,\\
{\cal O}^{5S}_{ijkl}  = (\bar{q}_{Li} H d_{Rj}) (\bar{\psi}_k \psi_l) \,,&\quad
{\cal O}^{\overline{5S}}_{ijkl} = (\bar{d}_{Ri} H^\dagger q_{Lj}) (\bar{\psi}_k \psi_l) = \left[{\cal O}^{5S}_{jilk}\right]^\dagger \,,\\
{\cal O}^{5P}_{ijkl} = (\bar{q}_{Li} H d_{Rj}) (\bar{\psi}_k i\gamma_5 \psi_l) \,,&\quad
{\cal O}^{\overline{5P}}_{ijkl} = (\bar{d}_{Ri} H^\dagger q_{Lj}) (\bar{\psi}_k i\gamma_5 \psi_l) = \left[{\cal O}^{5P}_{jilk}\right]^\dagger \,.
\end{align}
For simplicity, we neglect tensor operators.
Hermiticity requires that
\be
\begin{split}
C_{ijkl}^{1V}  =\left[C_{jilk}^{1V}\right]^*  \,,&\,
C_{ijkl}^{1A}  =\left[C_{jilk}^{1A}\right]^* \,,\,
C_{ijkl}^{2V}  =\left[C_{jilk}^{2V}\right]^*  \,,\,
C_{ijkl}^{2A}  =\left[C_{jilk}^{2A}\right]^* \,,\\
C_{ijkl}^{3V}  =\left[C_{jilk}^{3V}\right]^*  \,,&\,
C_{ijkl}^{3A}  =\left[C_{jilk}^{3A}\right]^* \,,\,
C_{ijkl}^{4S} =\left[C_{jilk}^{\overline{4S}}\right]^* \,,\,
C_{ijkl}^{4P} =\left[C_{jilk}^{\overline{4P}}\right]^* \,,\\
C_{ijkl}^{5S} =\left[C_{jilk}^{\overline{5S}}\right]^* \,,&\,
C_{ijkl}^{5P} =\left[C_{jilk}^{\overline{5P}}\right]^* \,.
\end{split}
\ee
Of these operators, ${\cal O}^{1V,1A}_{ijkl}$, ${\cal O}^{3V,3A}_{ijkl}$ and ${\cal O}^{5S,5P}_{ijkl}$ provide flavor violating interactions to the down quarks at the tree level, so that hereafter we only consider these three operators.

\subsubsection{$\psi\sim(\3,\1,\1)$}

The leading coefficients are given by 
\begin{align}
\label{eq:cF311_1}
C_{ijkl}^{1V,1A} & = c^{1V,1A}_1 \delta_{ij} \delta_{kl} + c^{1V,1A}_2 \delta_{il} \delta_{jk} + \dots \ , \\
\label{eq:cF311_3}
C_{ijkl}^{3V,3A} & = c^{3V,3A}_1 \delta_{ij} \delta_{kl} + \dots \ , \\
\label{eq:cF311_5}
C_{ijkl}^{5S,5P} & = c^{5S,5P}_1 (Y_d)_{ij} \delta_{kl} + c^{5S,5P}_2 (Y_d)_{kj} \delta_{il} + \dots \ ,
\end{align}
where the ellipses indicate further Yukawa matrix insertions and $c^n_m$ are taken to be real.

Again, we take $(Y_d)_{ij}=y_u^i \delta_{ij}$ and $Y_u=V^\dagger \hat{Y}_u$ without loss of generality.
In the unitary gauge $\vev{H}=(0,v/\sqrt{2})^T$ and 
in the quark mass basis ($u_L \to V^\dagger u_L$), we find 
\begin{align}
\L^{\eff}_F \supset 
\frac{1}{\Lambda^2} \sum_{i,j}\Bigg\{& 
    c^{1V}_2 (\bar{d}_{Li} \gamma^\mu d_{Lj})(\bar\psi_j \gamma_\mu \psi_i)
   + c^{1V}_2 \sum_{k,l} V_{il} V^*_{jk}( \bar{u}_{Li} \gamma^\mu u_{Lj}) (\bar\psi_k \gamma_{\mu} \psi_l) \nonumber \\
   & + c^{1A}_2 (\bar{d}_{Li} \gamma^\mu d_{Lj})(\bar\psi_j \gamma_\mu \gamma_5 \psi_i)
   + c^{1A}_2 \sum_{k,l} V_{il} V^*_{jk}( \bar{u}_{Li} \gamma^\mu u_{Lj}) (\bar\psi_k \gamma_{\mu}\gamma_5 \psi_l)\Bigg\} \nonumber \\
    +\frac{1}{\Lambda^3} \sum_{i,j}\Bigg\{& c^{5S}_2 (m_d^j+m_d^i) (\bar{d}_{i} d_{j}) (\bar\psi_j \psi_i)+ c^{5S}_2 (m_d^j-m_d^i) (\bar{d}_{i} \gamma_5 d_{j}) (\bar\psi_j \psi_i) \, , \nn\\
    \label{eq:LF311}
   & + c^{5P}_2 (m_d^j+m_d^i) (\bar{d}_{i} d_{j}) (\bar\psi_j i\gamma_5 \psi_i)+ c^{5P}_2 (m_d^j-m_d^i) (\bar{d}_{i} \gamma_5 d_{j}) (\bar\psi_j i\gamma_5 \psi_i) \Bigg\} \, ,
\end{align}
where $m_d^i = y_d^i v /\sqrt{2}$.
Note that in the last equation both quarks and DM fields are in the mass basis.

\subsubsection{$\psi\sim(\1,\3,\1)$}

The leading coefficients are given by 
\begin{align}
\label{eq:cF131_1}
C_{ijkl}^{1V,1A} & = c^{1V,1A}_1 \delta_{ij} \delta_{kl} + \dots \ , \\
\label{eq:cF131_3}
C_{ijkl}^{3V,3A} & = c^{3V,3A}_1 \delta_{ij} \delta_{kl} + \dots \ , \\
\label{eq:cF131_5}
C_{ijkl}^{5S,5P} & = c^{5S,5P}_1 (Y_d)_{ij} + \dots \ ,
\end{align}
where the ellipses indicate further Yukawa matrix insertions. 
No flavor violating interaction in the down sector exists, and therefore $\psi\sim(1,3,1)$ is not a suitable representation to address the Belle~II and NA62 excesses.

\subsubsection{$\psi\sim(\1,\1,\3)$}

The leading coefficients are given by 
\begin{align}
\label{eq:cF113_1}
C_{ijkl}^{1V,1A} & = c^{1V,1A}_1 \delta_{ij} \delta_{kl} + \dots \ , \\
\label{eq:cF113_3}
C_{ijkl}^3 & = c^{3V,3A}_1 \delta_{ij} \delta_{kl} + c^{3V,3A}_2 \delta_{il} \delta_{jk} + \dots \ ,\\
\label{eq:cF113_5}
C_{ijkl}^{5S,5P} & = c^{5S,5P}_1 (Y_d)_{ij} \delta_{kl} + c^5_2 (Y_d)_{il} \delta_{jk} + \dots \ ,
\end{align}
where the ellipses indicate further Yukawa matrix insertions and $c^n_m$ are taken to be real. 
The flavor violating interactions arise only from ${\cal O}^{3V,3A}_{ijkl}$ and ${\cal O}^{5S,5P}_{ijkl}$, which are of the form
\begin{align}
\L^{\eff}_F \simeq \frac{1}{\Lambda^2} \sum_{i,j}\Bigg\{& c^{3V}_2 \(\bar{d}_{Li} \gamma^\mu d_{Lj} \) (\bar\psi_j \gamma_{\mu} \psi_i)
+c^{3A}_2 \(\bar{d}_{Li} \gamma^\mu\gamma_5 d_{Lj} \) (\bar\psi_j \gamma_{\mu} \psi_i) \Bigg\}
\nn \\ 
+\frac{1}{\Lambda^3} \sum_{i,j}\Bigg\{& c^{5S}_2 (m_d^i+m_d^j) (\bar{d}_{i} d_{j}) (\bar\psi_j \psi_i)
+ c^{5P}_2 (m_d^i+m_d^j) (\bar{d}_{i} d_{j}) (\bar\psi_j i\gamma_5\psi_i)
\nn \\ 
&+ c^{5S}_2 (m_d^i-m_d^j) (\bar{d}_{i} \gamma_5 d_{j}) (\bar\psi_j \psi_i)
+ c^{5P}_2 (m_d^i-m_d^j) (\bar{d}_{i} \gamma_5 d_{j}) (\bar\psi_j i\gamma_5 \psi_i)
\Bigg\} \, ,
\end{align}
where we take the unitary gauge and the quark mass basis. Note that the $d_i\to d_j$ operators of this Lagrangian are identical to those of Eq.~\eqref{eq:LF311} with $c^{1V,1A}_2\to c^{3V,3A}_2$.

\section{Experimental searches for $d_i\to d_j + \slashed{E}$}
\label{sec:exp}

The framework of MFV DM leads to a prediction of a missing energy signature in FCNC rare meson decays with flavor correlations given by the MFV structure.

Recently, interesting excess of events has been observed both in $B^+\to K^+ \nu\bar\nu$ at Belle~II~\cite{Belle-II:2023esi} and in $K^+\to\pi^+\nu\bar\nu$ at NA62~\cite{NA62:2024pjp}.

The Belle~II excess over the SM prediction~\cite{Parrott:2022zte,Becirevic:2023aov} has a statistical significance of $2.7\sigma$~\cite{Belle-II:2023esi} and has triggered extensive phenomenological investigations \cite{Athron:2023hmz,Bause:2023mfe,Allwicher:2023xba,Felkl:2023ayn,Wang:2023trd,He:2023bnk,Chen:2023wpb,Berezhnoy:2023rxx,Datta:2023iln,Altmannshofer:2023hkn,McKeen:2023uzo,Fridell:2023ssf,Ho:2024cwk,Loparco:2024olo,Gabrielli:2024wys,Chen:2024jlj,Hou:2024vyw,Chen:2024cll,He:2024iju,Bolton:2024egx,DAlise:2024qmp,Marzocca:2024hua,Becirevic:2024pni,Buras:2024ewl,Kim:2024tsm,Rosauro-Alcaraz:2024mvx,Hati:2024ppg,Zhang:2024rmb,Chen:2025npb,Crivellin:2025qsq,Allwicher:2024ncl,Becirevic:2024iyi,Altmannshofer:2024kxb,Guedes:2024vuf,Buras:2024mnq,Bhattacharya:2024clv,Hu:2024mgf,Lee:2025jky,Lin:2025jzp,Calibbi:2025rpx,Berezhnoy:2025tiw,Bolton:2025fsq,Aliev:2025hyp,Ding:2025eqq,Abumusabh:2025zsr,Shaw:2025ays,Abdughani:2023dlr,He:2025jfc,Ko:2025drr,Berezhnoy:2025nmb,Kim:2025zaf,Li:2025ski,Gao:2025ohi,Alda:2025uwo,DiLuzio:2025qkc,Bolton:2025lnb}. In particular, Ref.~\cite{Bolton:2024egx} considers a scenario where the excess is due to the emission of single dark state coupled to the SM fields using an EFT description.

The most recent NA62 measurement of the $K^+\to\pi^+\nu\bar\nu$ decay~\cite{NA62:2024pjp} uses data from 2016--2022 observing $51$ candidate events and a total background of $18^{+3}_{-2}$, representing the first observation of the decay at a significance in excess of $5\sigma$.
The branching ratio is measured to be $\mathcal{B}(K^{+}\rightarrow\pi^{+}\nu\bar{\nu}) = \left(13.0^{+3.3}_{-3.0}\right)\times10^{-11}$, this can be compared to SM predictions of around $8\times10^{-11}$ with a precision of better than $10\%$~\cite{Buras:2022wpw,Anzivino:2023bhp,DAmbrosio:2022kvb}.
Therefore, the central value is around $50\%$ larger than expected, although the current experimental uncertainty is such that this corresponds to an excess with significance of around $1.7\sigma$.

\begin{table}[t!]
\centering
\begin{tabular}{c|c|c}
Flavor transition & Meson decay & Experiment \\
\midrule
\midrule
\multirow{4}{*}{$b \to s$} 
  & $B^+ \to K^+ +\slashed{E}$ & Belle~II~\cite{Belle-II:2023esi}, BaBar~\cite{BaBar:2013npw} \\ 
  & $B^0 \to K^0 +\slashed{E}$ & BaBar~\cite{BaBar:2013npw} \\
  & $B^+ \to K^{*+} +\slashed{E}$ & BaBar~\cite{BaBar:2013npw} \\
  & $B^0 \to K^{*0} +\slashed{E}$ & BaBar~\cite{BaBar:2013npw} \\
\midrule
\midrule
\multirow{4}{*}{$b \to d$} 
  & $B^+ \to \pi^+ +\slashed{E}$ & Belle~\cite{Belle:2013tnz} \\ 
  & $B^0 \to \pi^0 +\slashed{E}$ & Belle~\cite{Belle:2013tnz} \\
  & $B^+ \to \rho^{+} +\slashed{E}$ & Belle~\cite{Belle:2013tnz} \\
  & $B^0 \to \rho^{0} +\slashed{E}$ & Belle~\cite{Belle:2013tnz} \\
\midrule
\midrule
\multirow{2}{*}{$s \to d$} 
  & $K^+ \to \pi^+ + \slashed{E}$ & NA62~\cite{NA62:2018ctf,NA62:2020fhy,NA62:2021zjw,NA62:2024pjp} \\ 
  & $K_L \to \pi^0 + \slashed{E}$ & KOTO~\cite{KOTO:2024zbl} \\
\end{tabular}
\caption{List of FCNC rare mesons with missing energy experiments for each flavor transition.}
\label{tab:mesondecays}
\end{table}

The main channels of rare meson decays for each $d_i\to d_j$ transition are shown in Table~\ref{tab:mesondecays}.

\subsection{NA62}

The small branching fraction of the $K^{+}\rightarrow\pi^{+}\nu\bar{\nu}$ decay in the SM~\cite{Buras:2022wpw,Anzivino:2023bhp,DAmbrosio:2022kvb}, which predicts that this process occurs fewer than once in ten billion $K^+$ decays, is dominated by short-distance contributions, and the necessary hadronic matrix element can be extracted from {\em data} using isospin. The decay is thus both uniquely rare and uniquely theoretically clean.

The NA62 experiment at CERN~\cite{NA62:2024pjp} uses a decay-in-flight technique and defines two signal regions as
\beq
m_\text{miss}^2 \in \begin{cases}
[0,0.01] \, \text{GeV}^2 &\text{ for Signal Region 1 (SR1),} \\
[0.026,0.068] \, \text{GeV}^2 &\text{ for Signal Region 2 (SR2),}
\end{cases}
\eeq
where $m_{\rm miss}^{2} = (P_{K} - P_{\pi})^{2} \equiv q^{2}$ and $P_{K}$ and $P_{\pi}$ are the 4-momenta of the kaon and pion, respectively.
The number of excepted events can be expressed as
\beq
\label{eq:sesN}
\frac{d N_\text{signal}}{dq^2}= \,\, \text{SES}^{-1}(q^2) \times \frac{d\BR}{dq^2} \,,
\eeq
where SES is the single event sensitivity, which contains all of the experimental details of the signal sensitivity.

Despite the small branching fraction of the $K^{+}\rightarrow\pi^{+}\nu\bar{\nu}$ decay mode~\cite{Buras:2022wpw,Anzivino:2023bhp,DAmbrosio:2022kvb}, NA62 has provided experimental evidence of this decay, with significance in excess of $5\sigma$, and obtained a central value slightly larger than the SM prediction.
In addition, the same 2016-2022 dataset has been interpreted in terms of a search for $K^{+}\rightarrow\pi^{+}X$, where $X$ is a new unobserved state. In this context, NA62 published the SES as a function of $m_{X} = m_{\rm miss}$, see the right panel of Fig.~1 in Ref.~\cite{NA62:2025upx}.
This can also be used, together with equation~\ref{eq:sesN}, to consider other BSM scenarios with $q^{2}$ dependence. Below, we will consider the case of MFV DM.

\subsection{KOTO}

The primary objective of the KOTO experiment at J-PARC is to search for the $K_{L}\rightarrow\pi^{0}\nu\bar{\nu}$ decay. Its branching ratio is predicted to be around $3\times10^{-11}$ in the SM~\cite{Buras:2022wpw,Anzivino:2023bhp,DAmbrosio:2022kvb}.
This rarity, and the lack of charged tracks, makes this a formidable experimental challenge.

The KOTO experiment defines a signal region that covers the range $m_\text{miss}^2\in [0,0.065]$ GeV$^2$.
The number of expected events is given as in Eq.~\eqref{eq:sesN}.
The SES of the KOTO $K_{L}\rightarrow\pi^{0}\nu\bar{\nu}$ analysis of the 2021 dataset is reported in~\cite{KOTO:2024zbl}. Zero events were observed in the signal region and upper limits were established at $\mathcal{B}(K_{L}\rightarrow\pi^{0}\nu\bar{\nu})<2.2\times10^{-9}$ at $90\%$ confidence level, as well as on $\mathcal{B}(K_{L}\rightarrow\pi^{0}X)$ for a set of mass hypotheses $m_{X}$, where $X$ is a hypothetical neutral state. 
Since no events were observed and the background is small, the SES as a function of $q^{2}$ is inferred by multiplying the $90\%$ confidence level $\mathcal{B}(K_{L}\rightarrow\pi^{0}X)$ limits by a factor of $2.3$.

\subsection{Belle~II}

Belle~II recently reported a measurement of $B^+ \to K^+ \nu\bar\nu$~\cite{Belle-II:2023esi} with central value notably above the SM prediction~\cite{Parrott:2022zte,Becirevic:2023aov,Buras:2014fpa,Brod:2010hi}.
For the $B^+ \to K^+ \nu\bar\nu$ analysis both the well-established
\emph{hadronic tagging analysis} (HTA) and the more recently developed
\emph{inclusive tagging analysis} (ITA)%
~\cite{Belle-II:2023esi,Belle:2019iji,Belle-II:2021rof} techniques were used, exploiting two almost independent datasets.

In the HTA, the complete decay chain of the tag-side $B$ meson is
reconstructed, which enables a direct determination of the di-neutrino
invariant mass $q^2$. On the other hand the ITA, while it benefits from
substantially larger statistics and therefore drives the overall
sensitivity, does not allow for a direct reconstruction of
$q^2$. Instead, the analysis relies on a proxy observable
$q^2_{\mathrm{rec}}$, defined as~\cite{Belle-II:2023esi}
\be
\label{eq:q2rec}
\begin{split}
q^2_{\rec}~\equiv~ & \frac{s}{4} + m_K^2 - \sqrt{s} E_K^* 
~=~q^2 + E_B^{*2} - m_B^2 - 2 \, \vec{p}_B^{\,*} \cdot \vec{p}_K^{\,*}(q^2) \\
~=~ & q^2 + E_B^{*2} - m_B^2
- 2 |\vec{p}_B^{\,*}|\gamma^{**}
\left( \beta^{**} E_K^{**}(q^2)
+ |\vec{p}_K^{\,**}(q^2)| \cos \theta^{**} \right)~,
\end{split}
\ee
where\footnote{$\lambda(x,y,z)=x^2 + y^2 + z^2 - 2xy -2yz -2zx$.}
$|\vec{p}_K^{\,**}| = \lambda^{1/2}(m_B^2,m_K^2,q^2)/(2 m_B)$ and
$\beta^{**} = |\vec{p}_B^{\,*}|/E_B^* = 0.06234$ denotes the boost
parameter relating the $B$-meson rest frame to the $B\bar B$ rest frame.
Quantities marked with a single (double) star are evaluated in the
$B\bar B$ ($B$) rest frame. Finally, $\theta^{**}$ is the polar angle of
the kaon momentum in the $B$-meson rest frame with respect to the
$B$-meson momentum in the $B\bar B$ rest frame, and is the only
unmeasured variable entering Eq.~\eqref{eq:q2rec}.

For fixed $q^2$, the distribution of $q^2_{\rec}$ follows from
Eq.~\eqref{eq:q2rec} by assuming a uniform distribution of
$c^{**}\equiv\cos\theta^{**}\in[-1,1]$. We neglect the sub-percent--level
uncertainties associated with the finite resolution on the energies and
momenta appearing in Eq.~\eqref{eq:q2rec}~\cite{BelleIITrackingGroup:2020hpx},
as their impact is subleading for the present analysis.

Accordingly, in the ITA the physical branching ratios are related to the
observed event yields as~\cite{Abumusabh:2025zsr}
\be
\label{eq:dNdq2rec}
\begin{split}
 \frac{d N_X}{d q^2_{\rec}} ~&=~
 N_B \int dq^2 \int d c^{**} \, \frac{1}{2} \,
 \delta\Bigl(q^2_{\rec} - q^2_{\rec}(q^2,c^{**})\Bigr) \,
 \eps_\text{ITA}(q^2) \,
 \frac{d \BR_X}{dq^2} \\
~&\equiv~ N_B \int dq^2 \,
 f_{q^2_{\rec}}(q^2) \,
 \eps_\text{ITA}(q^2) \,
 \frac{d \BR_X}{dq^2}~,
\end{split}
\ee
where $N_B = 387(6) \times 10^6$~\cite{Belle-II:2023esi} denotes the number
of $B$-meson candidates from decay channel $X$, $\BR$ is the
corresponding branching ratio, and $\eps_\text{ITA}(q^2)$ is the ITA
selection efficiency as a function of $q^2$, taken from
Ref.~\cite{Belle-II:2023esi}. The function $q^2_{\rec}(q^2,c^{**})$
is given by Eq.~\eqref{eq:q2rec}. 
The final expression introduces the smearing
function $f_{q^2_{\rec}}$ commonly used in the
literature~\cite{Fridell:2023ssf,Bolton:2024egx,Bolton:2025fsq}.

Within the HTA analysis Belle~II reconstructs $q^2$ directly on
an event-by-event basis. Therefore, the relation between the physical branching
ratios and the observed yields therefore takes the simpler form
\be
\label{eq:dNdq2}
 \frac{d N_X}{d q^2} ~=~
 N_B \, \eps_\text{HTA}(q^2) \,
 \frac{d \BR_X}{dq^2}~,
\ee
where $\eps_\text{HTA}(q^2)$ denotes the HTA selection efficiency as a
function of $q^2$, also provided in Ref.~\cite{Belle-II:2023esi}.

\subsection{Belle}

The Belle collaboration obtained upper limits at $90\%$ confidence level (CL) on $b\to d\nu\bar\nu$ decays~\cite{Belle:2017oht}, which remain the most stringent reported to date. Collectively,
\begin{align}
\label{eq:bellelimit1}
\mathcal{B}(B^{+}\rightarrow \rho^{+}\nu\bar{\nu}) &< 3.0\times10^{-5} \ , \\
\label{eq:bellelimit2}
\mathcal{B}(B^{+}\rightarrow\pi^{+}\nu\bar{\nu}) &< 1.5\times10^{-5} \ , \\
\label{eq:bellelimit3}
\mathcal{B}(B^{0}\rightarrow\rho^{0}\nu\bar{\nu}) &< 4.0\times10^{-5} \ , \\
\label{eq:bellelimit4}
\mathcal{B}(B^{0}\rightarrow\pi^{0}\nu\bar{\nu}) &< 9.0\times10^{-4} \ .
\end{align}
In this case, the full event kinematics is not reconstructed, and thus the $q^{2}$ dependence is not available. 
Therefore, we use these global upper limits to constrain our models.
Future investigations of these decay modes may be possible at Belle-II.

\subsection{BaBar}

BaBar searched for $B\rightarrow K^{(*)}\nu\bar{\nu}$ decays~\cite{BaBar:2013npw} using a sample of 
$5\times10^{8}$ $B\bar{B}$ pairs from 
$\Upsilon(4s)\rightarrow B\bar{B}$ decays at the PEP-II facility, SLAC. 
In each event a hadronic decay of one of the $B\bar{B}$ pair is fully reconstructed and the $B\rightarrow K^{(*)}\nu\bar{\nu}$ search is performed in the remainder of the event, which is required to have missing energy, and therefore reconstructed $q^{2}>0$. 
Four channels were constrained,
with upper limits on branching ratios established at $90\%$ confidence level of~\cite{BaBar:2013npw}
\begin{align}
    \mathcal{B}(B^{+}\rightarrow K^{+}\nu\bar{\nu}) < 1.6 \times10^{-5}\,\,,
    \\
    \mathcal{B}(B^{0}\rightarrow K^{*0}\nu\bar{\nu}) < 4.9 \times10^{-4}\,\,,
    \\
    \mathcal{B}(B^{+}\rightarrow K^{*+}\nu\bar{\nu}) < 6.4 \times10^{-5}\,\,,
    \\
    \mathcal{B}(B^{0}\rightarrow K^{*0}\nu\bar{\nu}) < 1.2 \times10^{-4}\,\,.
\end{align}

\section{Results}
\label{sec:results}

In this section we present the results of our analysis. 
In each case, we evaluate the improvement of our model's description compared to the SM. 
To do so, we construct a likelihood $\mcL_\text{BSM}(\boldsymbol{\theta},\boldsymbol{c})$ as described in Appendix~\ref{app:likelihood}, where $\boldsymbol{\theta}$ and $\boldsymbol{c}$ are the set of nuisance parameters and BSM couplings, respectively. The SM likelihood is obtained by setting $\boldsymbol{c}=0$, i.e. $\mcL_\text{SM}(\boldsymbol{\theta})=\mcL_\text{BSM}(\boldsymbol{\theta},\boldsymbol{c}=0)$.
The improvement is then measured by the test statistic
\be
\Delta\equiv -2\ln\frac{\hat\mcL_\text{BSM}}{\hat\mcL_\text{SM}} \ ,
\ee
where a hat indicates the value of a likelihood at the best-fit point of the data.
Negative values of $\Delta$ indicate that the BSM hypothesis provides a better fit to the data than the SM~\footnote{When the usual regularity conditions are satisfied,  $-\Delta= \chi^2_{\rm BSM} - \chi^2_{\rm SM}$ is asymptotically distributed as a chi-square variable with a number of degrees of freedom equal to the number of additional parameters introduced by the BSM hypothesis:
so that, for instance, for a single added BSM parameter $-\Delta = 1$ corresponds roughly to a $1\sigma$ preference, 
$-\Delta = 4$ to $2\sigma$, $-\Delta = 9$ to $3\sigma$, and so on.}.

\subsection{Scalar DM}
\label{sec:Results_ScalarDM}

\begin{figure}[t!]
\begin{center}
\includegraphics[scale=0.6]{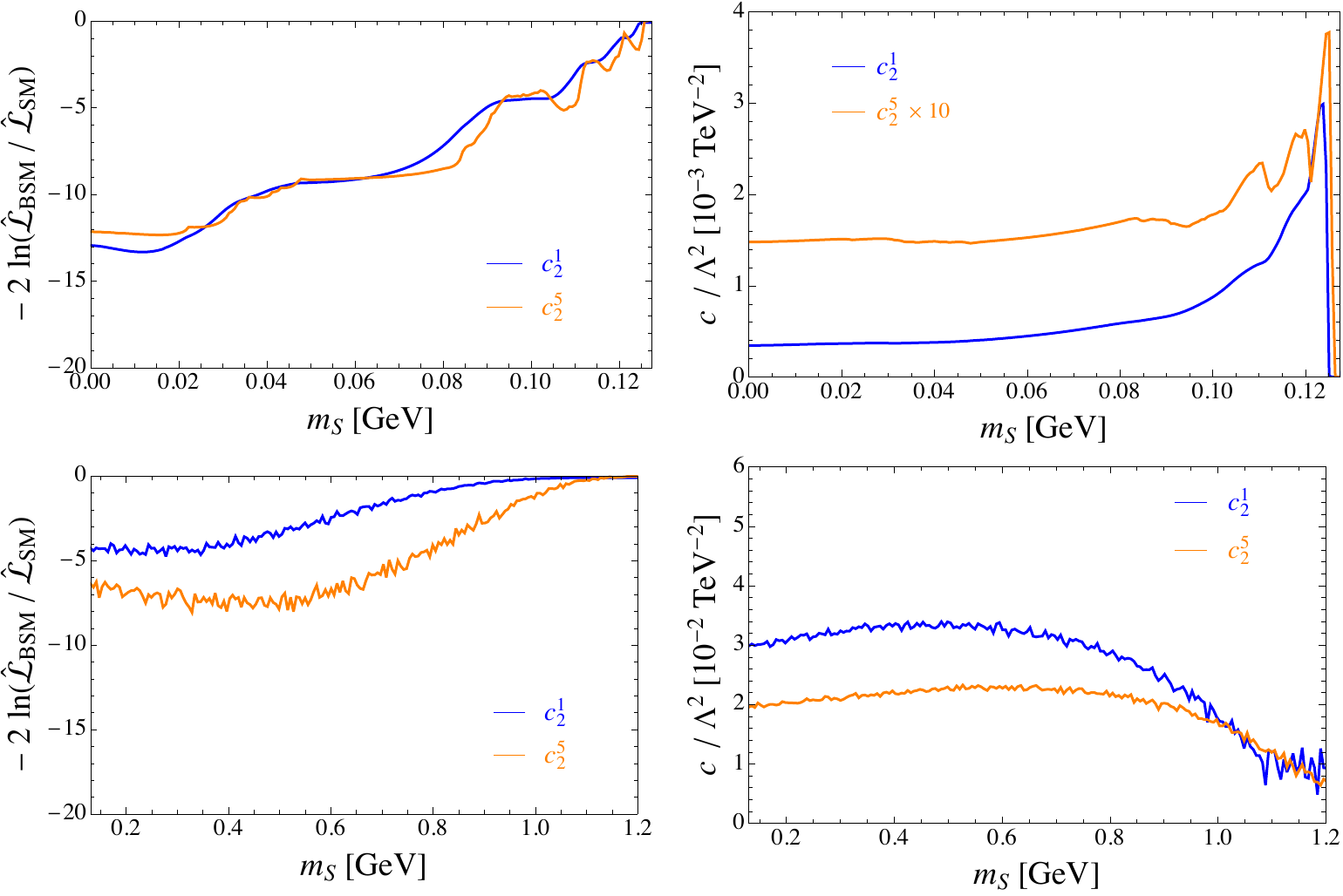}
\caption{Results for the case of scalar DM $S\sim(3,1,1)$.
Left: test statistic as a function of the scalar mass, $m_{S}$.
Right: best-fit value of new physics parameter, $\Lambda$, as a function of the scalar mass.
}
\label{fig:S311}
\end{center}
\end{figure}

We first focus on the case of scalar DM transforming as $S\sim(3,1,1)$. The set of flavor-violating couplings for this case is given by $c_2^1$ and $c_2^5$ from Eqs.~\eqref{eq:cS311_1}--\eqref{eq:cS311_5}. We will study the improvement $\Delta$ of this model compared to the SM case with one flavor-violating coupling turned on at a time.

The DM modes of the meson decays are calculated from the Lagrangian Eq.~\eqref{eq:LS311} with the formulae provided in Appendix~\ref{app:rates} and the form factors taken as in Appendix~\ref{app:FF}.

Our findings are shown in Fig.~\ref{fig:S311}.
In the upper panels, we show the values of the improvement $\Delta$ (left) and the best-fit values of the BSM coefficient (right) for values of the DM mass within the mass range $0$--$125\,\text{MeV}$ covered by the NA62 and KOTO studies of $K\rightarrow\pi\nu\bar{\nu}$ decays.
Within this range, the model explains the NA62 excess, with this scenario favored by up to approximately $3\sigma$ at low masses. 
However, a description of the Belle~II excess is found to require larger coupling values which are constrained by the NA62 measurements. 
This means the Belle~II result is not simultaneously explained satisfactorily by a scalar DM particle of such a low mass.
When the DM mass is large enough, $\gtrsim 150\,\text{MeV}$, to avoid the constraint from NA62 and KOTO studies of $K\to \pi \nu\nu$ decays, we get the results in the lower panels for the values of the improvement $\Delta$ (left) and the best-fit values of the BSM coefficient (right).
In this scenario, driven by experimental results from the Belle~II excess, this BSM scenario provides an improved description over the SM, favored by up to $2$--$3\sigma$ at $m_{S}\sim0.5\,\text{GeV}$. 

The results for the $S\sim(1,1,3)$ case are qualitatively the same to this case and so are not shown here explicitly.

In conclusion, we find that a simultaneous description of the data, which is driven by the potential excesses from NA62 and Belle~II, is not readily provided by this MFV scenario with a scalar DM candidate. 
To provide such an improved simultaneous description would require a hierarchical coupling structure beyond the MFV.

\subsection{Fermion DM}
\label{sec:Results_FerionDM}

\begin{figure}[t!]
\begin{center}
\includegraphics[scale=0.6]{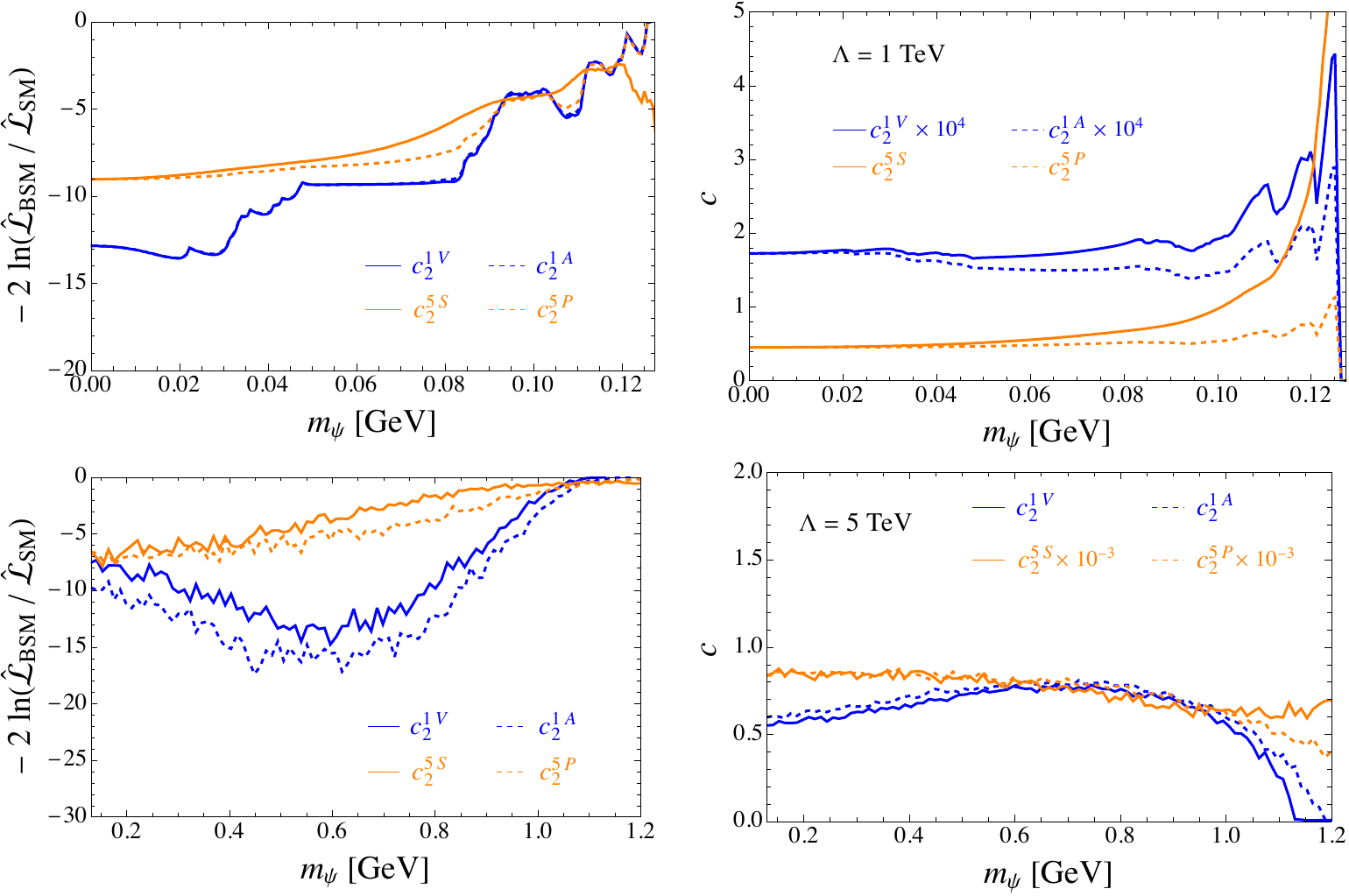}
\caption{Results for the case of fermion DM $\psi\sim(3,1,1)$.
Left: test statistic as a function of the fermion mass, $m_{\psi}$.
Right: best-fit value of new physics parameter, $\Lambda$, as a function of the fermion mass.
}
\label{fig:F311}
\end{center}
\end{figure}

We now consider the case of fermion DM transforming as $\psi\sim(3,1,1)$. 
The set of flavor-violating couplings for this case is given by $c_2^{1V}$, $c_2^{1A}$, $c_2^{5S}$ and $c_2^{5P}$ from Eqs.~\eqref{eq:cF311_1}--\eqref{eq:cF311_5}. 
We will study the improvement $\Delta$ of this model compared to the SM case with one flavor-violating coupling turned on at a time.

The DM modes of the meson decays are calculated from the Lagrangian Eq.~\eqref{eq:LF311} with the formulae provided in Appendix~\ref{app:rates} and the form factors taken as in Appendix~\ref{app:FF}.

Our results are displayed in Fig.~\ref{fig:F311}.
In the upper panels, we show the values of the improvement $\Delta$ (left) and the best-fit values of the BSM coefficient (right) for values of the DM
mass range $0$--$125\,\text{MeV}$ covered by the NA62 and KOTO studies of $K\rightarrow\pi\nu\bar{\nu}$ decays.
We observe the same behavior as in the scalar DM case:
the NA62 excess can be accommodated within this low mass range (with our model improving over the BSM description by up to $3\sigma$), but the Belle~II excess requires larger coupling values which are constrained by the NA62 measurements.
When the DM mass is large enough to avoid the constraint from NA62 and KOTO, and thus the excess at NA62 is left unexplained, we get the results in the lower panels for the values of the improvement $\Delta$ (left) and the best-fit values of the BSM coefficient (right). Here our MFV model with a DM fermion provides a description favored by $\sim2$--$3\sigma$ (at $m_{\psi}\sim0.5\,\text{GeV}$ over that provided by the SM).

We find results for the $\psi\sim(1,1,3)$ case are qualitatively the same, and therefore details are not presented explicitly.

Therefore, the qualitative conclusions are therefore the same as in section~\ref{sec:Results_ScalarDM}: while improved descriptions can be found individually for potential $K\rightarrow\pi\nu\bar{\nu}$ and $B\rightarrow K\nu\bar{\nu}$ excesses, a satisfactory simultaneous description has not been found in this framework.

\section{Conclusions}
\label{sec:conclusion}

Current rare-decay programs, in particular at NA62 and Belle~II, provide an exceptional window into missing-energy signatures potentially linked to dark matter emission. 
Within this context, we have investigated flavored dark matter in the MFV framework, focusing on the processes $ d_i \to d_j + \slashed{E}$.
The MFV hypothesis offers a theoretically robust and predictive structure for incorporating new sources of flavor violation and naturally stabilizes the 
lightest component of a new flavored multiplet, making it an appealing setting for dark matter model building.

Our results show that the MFV setup can account for the observed excesses in either $K^+ \to \pi^+ \nu\bar{\nu}$ or $B^+ \to K^+ \nu\bar{\nu}$ individually, providing descriptions that are favored with respect to the SM by $\sim3\sigma$. However, a satisfactory simultaneous explanation of both channels cannot be achieved within a minimal construction containing only a single dark matter multiplet with degenerate components. 
Reconciling both excesses may require either introducing an additional multiplet with a different mass for each flavor, or generating sizeable mass splittings within a single multiplet. 
While such extensions remain fully consistent with MFV, they move away from strict minimality and may appear less economical from a theoretical point of view. Therefore, we do not pursue these non‑minimal possibilities further in this work.

Despite these structural limitations, our findings highlight the rich interplay between flavored dark matter, MFV-based model building, and precision flavor experiments. 
They demonstrate that flavored dark matter remains a compelling and testable paradigm, capable of addressing both collider-scale anomalies and the cosmological dark matter abundance. 

Our analysis relies on a likelihood model built from available experimental data on  $d_i \to d_j + \slashed{E}$ channels, revisiting old datasets and incorporating the latest results to provide a comprehensive and up-to-date phenomenological assessment of the MFV DM scenario.

\section*{Acknowledgments}

We thank Ludovico Vittorio for useful discussions at the early stage of this work and Jernej Kamenik for providing further insights on the analysis of Ref.~\cite{Bolton:2024egx}.
The work of C.T. has received funding from the French ANR, under contracts ANR-19-CE31-0016 (`GammaRare') and ANR-23-CE31-0018 (`InvISYble'), that he gratefully acknowledges.
The work of S.O. is supported by JSPS KAKENHI Grant No.~JP25K17401 and by an appointment to the JRG Program at the APCTP through the Science and Technology Promotion Fund and Lottery Fund of the Korean Government and by the Korean Local Governments -- Gyeongsangbuk-do Province and Pohang City. The work of FM is supported by the European Union-Next Generation EU and by the Italian Ministry of University and Research (MUR) via the PRIN 2022 project No. 2022K4B58X-AxionOrigins.
This article is based upon work from COST Action COSMIC WISPers CA21106, supported by COST (European Cooperation in Science and Technology). 
We also wish to thank the Theory Group of the Laboratori Nazionali di Frascati for their hospitality during the first stage of this work.

\appendix

\section{Likelihood function}
\label{app:likelihood}

In this section we present the details of the likelihood constructed for the analysis of this work.
The total likelihood is given by the product of the likelihood of each experimental measurement times a likelihood term for the SM prediction of the di-neutrino modes, i.e.
\be
\mcL = \mcL_\text{Belle~II} \ \mcL_\text{BaBar} \ \mcL_\text{Belle} \ \mcL_\text{NA62} \ \mcL_\text{KOTO} \ \mcL_{\nu\bar\nu} \ .
\ee
Each term is presented and explained below.

\subsection{Belle~II}

We construct the likelihood term for the Belle~II measurement based on the public data of Fig. 17 (ITA) and Fig. 20 (HTA) from Ref.~\cite{Belle-II:2023esi}, similarly to Ref.~\cite{Abumusabh:2025zsr}. The likelihood function is constructed as the product of two contributions, i.e.
\be
{\mcL}_\text{Belle~II}={\mcL}_\text{Belle~II}^{(1)} {\mcL}_\text{Belle~II}^{(2)} \ .
\ee

The first and main term in the likelihood function represents a Poisson distribution $P$ for each bin, labeled by an index $i$. Assuming the bins to be uncorrelated, we get
\be
\mcL_\text{Belle~II}^{(1)} = \prod_i^{n_\bin} P(n^i_\obs; \lambda^i) \ .
\ee
Here $n^i_\obs$ is the number of observed events in bin $i$, defined by $q^2 \in [q^2_i, q^2_{i+1})$, and $\lambda^i$ is the corresponding expected number of events
\be
\label{eq:lai}
\begin{split}
\lambda^i ~=~  n^i_\text{DM} +  n^i_{\nu\bar\nu} 
+ (1+\sigma^\text{bkg} \, \theta) \, n^i_\text{bkg}~.
\end{split}
\ee
This expression applies to either the ITA or HTA dataset, where $n^i_\text{DM}$, $n^i_{\nu\bar\nu}$, and $n^i_\bkg$ are the yields in the $i^\th$ bin for the signal, the $B^+ \to K^+ \nu\bar\nu$ decay, and the background, respectively. Within the two methods, the first two components on the right hand side of Eq.~\eqref{eq:lai} are defined as
\begin{align}
\label{eq:nivv_ITA}
n^i_{X,\text{ITA}} =&N_B \int\limits_\text{$i^\text{th}$ bin} dq^2_{\rec} \int dq^2 \, f_{q^2_{\rec}}(q^2) \, \eps_\text{ITA}(q^2) \, \frac{d\BR_{X}}{dq^2}~,\\
\label{eq:nivv_HTA}
n^i_{X,\text{HTA}} =&N_B \int\limits_\text{$i^\text{th}$ bin} dq^2 \, \eps_\text{HTA}(q^2) \, \frac{d\BR_{X}}{dq^2}~.
\end{align}

In order to get an estimate of the uncertainty on the total background normalisation for ITA and HTA, we identify from Figs.~17 and 20 of Ref.~\cite{Belle-II:2023esi}
\be
\text{pull} = \frac{|n^i_{\rm dat} - n^i_{\rm hst}| }{ \sqrt{(\sigma^i_{\rm dat})^2 + (\sigma^i_{\rm hst})^2}} \ ,
\ee
where the subscripts `dat' and `hst' denote respectively the data and the value of the stacked histograms, and $\sigma^i_{\rm dat}$ the uncertainty in the data point. By inverting the pull distribution, we get $\sigma_\text{ITA}^\text{bkg}\approx1\%$ and $\sigma_\text{HTA}^\text{bkg}\approx13\%$.
Thus, we introduced two nuisance parameters, i.e. $\theta_\text{ITA}$ and $\theta_\text{HTA}$, both constrained by a normal Gaussian distribution in the likelihood as
\be
{\mcL}_\text{Belle~II}^{(2)} = \frac{1}{\sqrt{2\pi}} e^{-\theta^2_\text{ITA}/2} \times \frac{1}{\sqrt{2\pi}} e^{-\theta^2_\text{HTA}/2} \, .
\ee

Ref.~\cite{Abumusabh:2025zsr} has shown that this procedure applied to the $B^+\to K^+ \nu\bar\nu$ analysis produces results in good agreement with those of Belle~II.

\subsection{BaBar}

We construct the likelihood term for the BaBar measurement based on the public data of Fig. 5 from Ref.~\cite{BaBar:2013npw}.
The likelihood function is constructed as the product of two contributions for each of the four $b\to s$ decay channels listed in Table~\ref{tab:mesondecays}, i.e.
\be
{\mcL}_\text{BaBar}=\prod_{k} \ {\mcL}_\text{BaBar,k}^{(1)} \ {\mcL}_\text{BaBar,k}^{(2)} \ ,
\ee
where the index $k$ labels the decay channels.

Similarly to the previous case, the first and main term in the likelihood function represents a Poisson distribution $P$ for each bin, labeled again by an index $i$. Assuming the bins to be uncorrelated, we get
\be
\mcL_\text{BaBar,k}^{(1)} = \prod_i^{n_\bin} P(n^i_\obs; \lambda^i) \ ,
\ee
where, as before, $n^i_\obs$ is the number of observed events in bin $i$, defined by $q^2 \in [q^2_i, q^2_{i+1})$, and $\lambda^i$ is the corresponding expected number of events
\be
\begin{split}
\lambda^i ~=~ (1+\sigma^\text{eff} \, \theta_{\text{eff}}) \, (n^i_\text{DM} +  n^i_{\nu\bar\nu}) 
+ (1+\sigma^\text{bkg} \, \theta_\text{bkg}) \, n^i_\text{bkg}~,
\end{split}
\ee
with this expression applying to either of the four dataset and $n^i_\text{DM}$, $n^i_{\nu\bar\nu}$, and $n^i_\bkg$ being the yields in the $i^\th$ bin for the signal, the neutrino decay, and the background, respectively. For each dataset, the first two components of the latter are defined as
\begin{align}
\label{eq:nivv_babar}
n^i_{X} =&N_B \int\limits_\text{$i^\text{th}$ bin} dq^2 \, \varepsilon(q^2) \, \frac{d\BR_{X}}{dq^2}~.
\end{align}
Here the efficiency $\varepsilon(q^2)$ is obtained by interpolating from values presented in Fig.~6 of Ref.~\cite{BaBar:2013npw}.

\begin{table}[t!]
\centering
\begin{tabular}{c|c|c}
Dataset & $\sigma^\text{eff}$ & $\sigma^\text{bkg}$ \\
\midrule
\midrule
$B^+\to K^+ + \slashed{E}$ & $0.07$ & $0.21$ \\
\midrule
$B^0\to K^0 + \slashed{E}$ & $0.11$ & $0.22$ \\
\midrule
$B^+\to K^{*+} + \slashed{E}$ & $0.10$ & $0.17$ \\
\midrule
$B^0\to K^{*0} + \slashed{E}$ & $0.20$ & $0.14$ \\
\end{tabular}
\caption{Estimates of the uncertainty on the total background and efficiency normalisation for each dataset of BaBar.}
\label{tab:sigmababar}
\end{table}

In order to get an estimate of the uncertainty on the total background and efficiency normalisation for each dataset, we 
consider the relative uncertainty on the total numbers of background events and signal efficiencies in the first three bins in Table~5 of Ref.~\cite{BaBar:2013npw}. Assuming that these relative uncertainties are constant over the bins, we set $\sigma^\text{eff}$ and $\sigma^\text{bkg}$ as in Table~\ref{tab:sigmababar} and we introduce two nuisance parameters, i.e. $\theta_\text{eff}$ and $\theta_\text{bkg}$, both constrained by a normal Gaussian distribution in the likelihood as
\be
{\mcL}_\text{BaBar,k}^{(2)} = \frac{1}{\sqrt{2\pi}} e^{-\theta^2_{\text{eff},k}/2} \times \frac{1}{\sqrt{2\pi}} e^{-\theta^2_{\text{bkg},k}/2} \, ,
\ee
for each dataset.

\subsection{Belle}

In the case of Belle search, we have no access to the kinematic distribution of the events and, additionally, their measurements only put upper limits on the decay rates.
Therefore we construct a likelihood term that constrains only the total branching ratio (BR) of the DM signal such as
\be
{\mcL}_\text{Belle}^{(2)} \propto \prod_k \text{exp}\left\{-\frac{1}{2}\left(\frac{\BR_\text{DM}}{\sigma_{\BR_\text{DM}}}\right)^2\right\}
\label{eqn:BelleLike}
\ee
where $k$ labels the decay modes and we estimate
\be
\sigma_{\BR_\text{DM}} \approx \frac{\BR_\text{exp} - \BR_{\nu\bar\nu}}{\sqrt{\chi^2(90\%,1\text{dof})}} \ ,
\ee
with $\BR_\text{exp}$ the experimental limit in Eqs.~\eqref{eq:bellelimit1}-\eqref{eq:bellelimit4}, $\BR_{\nu\bar\nu}$ the SM prediction of the di-neutrino channel and $\chi^2(90\%,1\text{dof})$ the value of a chi-squared variable at 90\%CL with 1 degree of freedom. 
The numerator quantifies the constraint on possible additional BSM contributions, while the denominator accounts for rescaling to a $68\%$ confidence interval, such that Eq.~\eqref{eqn:BelleLike} is a Gaussian constraint term.

From Ref.~\cite{Bause:2021cna}, the SM predictions of the neutrino decay modes are
\begin{align}
\BR(B^+\to \pi^+ \nu\bar\nu) &= 1.2\times10^{-7} \ , \\
\BR(B^0\to \pi^0 \nu\bar\nu) &= 5.6\times10^{-8} \ , \\
\BR(B^+\to \rho^+ \nu\bar\nu) &= 4.9\times10^{-7} \ , \\
\BR(B^0\to \rho^0 \nu\bar\nu) &= 2.3\times10^{-7} \ .
\end{align}

\subsection{NA62}

To construct the likelihood for NA62, we follow a variation of the procedure of Ref.~\cite{NA62:2020xlg} similarly to Ref.~\cite{Guadagnoli:2025xnt}.
We perform a fully frequentist hypothesis test using a shape analysis on the invisible invariant mass, and an unbinned profile likelihood ratio test statistic, based solely on public data. The likelihood function is constructed as the product of three contributions, i.e.
\be
\mcL_\text{NA62}={\cal L}_\text{NA62}^{(1)} \ {\cal L}_\text{NA62}^{(2)} \ {\cal L}_\text{NA62}^{(3)} \ .
\ee

The first term is a Poisson distribution for the total number of events $n_\text{tot}$ with mean value equal to the total number of observed events, which reads
\be
\mcL_\text{NA62}^{(1)} = P(n_\obs; n_\text{tot}) \ .
\ee
As reported in Ref.~\cite{NA62:2024pjp}, NA62 observed a total number of events equal to $n_\text{obs}=51$ in the full 2016-2022 dataset.
The variable $n_\text{tot}$ can be decomposed as $n_\text{tot}=n_b+n_{\nu\bar\nu}+n_\text{DM}$, which are the number of background, neutrino and DM events, respectively.
The numbers of events $n_{\nu\bar\nu}$ and $n_{\text{DM}}$ are obtained from the BR by integrating Eq.~\eqref{eq:sesN} over the signal region.

Given $n_\text{tot}$, its components are distributed as a function of the invisible invariant mass squared $m_{\inv}^2$ according to a multinomial likelihood, yielding ${\cal L}_\text{NA62}^{(2)}$ as
\be
{\cal L}_\text{NA62}^{(2)}=
\prod_{j=1}^{n_\text{obs}} \left[ \frac{n_b}{n_\text{tot}}g_b (m_j^2) + \frac{n_{\nu\bar\nu}}{n_\text{tot}}g_{\nu\bar\nu} (m_j^2) + \frac{n_\text{DM}}{n_\text{tot}}g_\text{DM} (m_j^2) \right] \ ,
\ee
where $g_{b}(m_\inv^2)$ is a function reproducing the distribution of the background events, and normalized to unit integral over the signal regions. We took this function from published data, i.e. from the right panel of Fig.~10 in Ref.~\cite{NA62:2024pjp}.
Conversely, $g_{\nu\bar\nu,\text{DM}}(m_\inv^2)$ is the BR shape normalized to unit integral over the signal regions.
Finally, the last likelihood factor is a Poisson-distributed constraint term for the number of background events
\be
\mcL_\text{NA62}^{(3)} =\frac{(\tau n_b)^{n_\text{off}}}{n_\text{off}!}e^{-\tau n_b} \ ,
\ee
with $\tau=\mu_b/\sigma_b^2$ and $n_\text{off}=(\mu_b/\sigma_b)^2$~\cite{Cousins_2008} obtained from the estimated mean value $\mu_b$ and uncertainty $\sigma_b$ of the number of background events. From Ref~\cite{NA62:2024pjp} we take $\mu_b=18.0$ and $\sigma_b=2.5$, with the latter obtained by averaging over the asymmetric uncertainty.

\subsection{KOTO}

The likelihood for the KOTO measurement is constructed similarly to the one for NA62. In this case, according to Ref.~\cite{KOTO:2024zbl}, we set $n_\text{obs}=0$, $\mu_b=0.252$ and $\sigma_b=0.08$, where the latter is obtained by averaging over the asymmetric statistical uncertainty and then adding in quadrature the statistical and systematic uncertainties.

\subsection{SM prediction of neutrino modes}

The general theoretical expressions of the rare meson decay rates with neutrino or DM emission can be found in Appendix~\ref{app:rates}.
These expressions depend on form factors which are introduced to parametrize the hadronic matrix elements of operators involving quark fields, see formulae in Appendix~\ref{app:FF}.
We will take the needed form factors from the most recent extrapolation based on lattice or experimental data available in literature, see Appendix~\ref{app:FF} for details.
In this work we will neglect the uncertainties related to the form factors of the decay rates, in practical terms this means that uncertainties on the expected shape of the $q^{2}$ distribution are neglected.

\begin{table}[t!]
\centering
\begin{tabular}{c|c|c}
$\BR_i$ & $\mu_{\BR_i}\pm\sigma_{\BR_i}$ & Ref. \\
\midrule
\midrule
$\BR(B^+\to K^+ \nu\bar\nu)$ & $(4.4\pm\sqrt{0.14^2+0.27^2})\times10^{-6}$ & \cite{Becirevic:2023aov} \\
\midrule
$\BR(B^0\to K^0 \nu\bar\nu)$ & $(2.05\pm\sqrt{0.07^2+0.12^2})\times10^{-6}$ & \cite{Becirevic:2023aov} \\
\midrule
$\BR(B^+\to K^{*+} \nu\bar\nu)$ & $(9.79\pm\sqrt{1.30^2+0.60^2})\times10^{-6}$ & \cite{Becirevic:2023aov} \\
\midrule
$\BR(B^0\to K^{*0} \nu\bar\nu)$ & $(9.05\pm\sqrt{1.25^2+0.55^2})\times10^{-6}$ & \cite{Becirevic:2023aov} \\
\midrule
\midrule
$\BR(K^+\to \pi^+ \nu\bar\nu)$ & $(8.4\pm1.0)\times10^{-11}$ & \cite{Buras:2014fpa} \\
\midrule
$\BR(K_L\to \pi^0 \nu\bar\nu)$ & $(3.4\pm0.6)\times10^{-11}$ & \cite{Buras:2014fpa} \\
\end{tabular}
\caption{SM predictions for the $B\to K^{(*)}\nu\bar\nu$ (without long distance contribution) and $K\to\pi\nu\bar\nu$'s BRs.}
\label{tab:BRneutrinos}
\end{table}

In contrast, we introduce a likelihood term to take into account the uncertainties on the SM prediction of the {\it total} BRs for the $B\to K^{(*)}\nu\bar\nu$ and $K\to\pi\nu\bar\nu$, which are taken from literature as reported in Table~\ref{tab:BRneutrinos}.
We hence define a nuisance parameter for each of them as 
\be
\theta_{\BR_i} \equiv \frac{\BR_i-\mu_{\BR_i}}{\sigma_{\BR_i}} \ ,
\ee
where $\mu_{\BR_i}$ and $\sigma_{\BR_i}$ are the central value and the uncertainty of the BR ${\BR_i}$.
Finally, we introduce a normal Gaussian distribution in the likelihood for each of them as
\be
{\mcL}_{\nu\bar\nu} = \prod_{\BR_i} \frac{1}{\sqrt{2\pi}} e^{-\theta^2_{\BR_i}/2} \, .
\ee
We do not introduce instead any additional term for the $B\to \pi\nu\bar\nu$ and $B\to\rho\nu\bar\nu$'s BRs as they are poorly constrained.

On a practical level, we found that the uncertainty on the SM predictions of the neutrino modes has negligible impact on our numerical results.

\section{Decay rates}\label{app:rates}
\label{sec:app_DecayRates}

In this appendix we report the theoretical formulae for the rate of FCNC rare meson decay with DM pair emission for the scalar and fermion case.

\subsection{$d_j\to d_i SS^\dagger$}

In the LEFT, the full set of operators up to dimension six is given by
\begin{equation}
\begin{aligned}
\mathcal{L}_\text{LEFT} \supset \ & g_{VV}^{ijkl}(\overline{d}_i\gamma_\mu d_j)(S^\dagger_k i \lrpartial{\mu} S_l)
+ g_{AV}^{ijkl}(\overline{d}_i\gamma_\mu\gamma_5 d_j)(S^\dagger_k i \lrpartial{\mu} S_l) \\
& + g_{SS}^{ijkl}(\overline{d}_i d_j)(S^\dagger_k S_l)
+ g_{PS}^{ijkl}(\overline{d}_i\gamma_5 d_j)(S^\dagger_k S_l) \ .
\end{aligned}
\end{equation}
From those operators, assuming degenerate DM masses, the meson decay rates are given by~\cite{Bolton:2024egx}
\begin{equation}
\begin{aligned}
\frac{d\Gamma(P(d_j)\to \tilde{P}(d_i) S_k S_l^\dagger)}{dq^2}=\frac{p_{\tilde{P}}}{96\pi^3} \sqrt{1-\frac{4m_S^2}{q^2}} \Bigg[&\frac{3}{4}|g_{SS}^{ijkl}|^2\frac{(m_{P}^2-m_{\tilde{P}}^2)^2}{m_{P}^2(m_{i}-m_{j})^2}f_0^2(q^2)\\
+&|g_{VV}^{ijkl}|^2 p_{\tilde{P}}^2 \left(1-\frac{4m_S^2}{q^2}\right) f_+^2(q^2)\Bigg] \ , \\
\frac{d\Gamma(P(d_j)\to V(d_i) S S^\dagger)}{dq^2}=\frac{p_{V}}{96\pi^3}\sqrt{1-\frac{4m_S^2}{q^2}}\Bigg[&3|g_{PS}^{ijkl}|^2\frac{p_{V}^2}{(m_{q_i}+m_{q_j})^2}A_0^2(q^2)\\
+2&|g_{VV}^{ijkl}|^2 \left(q^2-4m_S^2\right) \frac{p_V^2}{(m_P+m_V)^2} V^2(q^2)\\
+|g_{AV}^{ijkl}|^2 \left(q^2-4m_S^2\right)& \left(\frac{(m_P+m_V)^2}{2m_P^2}A_1^2(q^2)+\frac{16m_V^2}{q^2}A_{12}^2(q^2)\right)\Bigg] \ ,
\end{aligned}
\end{equation}
where $P(\tilde{P})$ and $V$ denotes respectively pseudoscalar and vector mesons.

For the decay of $K_L\to\pi^0$, one must substitute
\begin{equation}
\begin{aligned}
g_{SS}^{ijkl} \ & \to \ \sqrt{2}\ \text{Re}\left(g_{SS}^{12kl}\right) \,,\quad
g_{VV}^{ijkl} \ & \to \ \sqrt{2}\ \text{Im}\left(g_{VV}^{12kl}\right) \,.
\end{aligned}
\end{equation}

\subsection{$d_j\to d_i \psi\overline{\psi}$}

In the LEFT, the full set of operators up to dimension six is given by
\begin{equation}
\begin{aligned}
\L_\text{LEFT} \supset \ \ \ &f_{VV}^{ijkl}(\overline{d}_i\gamma_\mu d_j)(\overline{\psi}_k  \gamma^\mu \psi_l)+f_{AV}^{ijkl}(\overline{d}_i\gamma_\mu\gamma_5 d_j)(\overline{\psi}_k  \gamma^\mu\gamma_5 \psi_l) \\
+&f_{VA}^{ijkl}(\overline{d}_i\gamma_\mu d_j)(\overline{\psi}_k  \gamma^\mu\gamma_5 \psi_l)+f_{AA}^{ijkl}(\overline{d}_i\gamma_\mu\gamma_5 d_j)(\overline{\psi}_k  \gamma^\mu\gamma_5 \psi_l) \\
+&f_{SS}^{ijkl}(\overline{d}_i d_j)(\overline{\psi}_k  \psi_l)+f_{PS}^{ijkl}(\overline{d}_i\gamma_5 d_j)(\overline{\psi}_k  \psi_l) \\
+&f_{SP}^{ijkl}(\overline{d}_i d_j)(\overline{\psi}_k\gamma_5  \psi_l)+f_{PP}^{ijkl}(\overline{d}_i\gamma_5 d_j)(\overline{\psi}_k \gamma_5 \psi_l) \ .
\end{aligned}
\end{equation}
From those operators, assuming degenerate DM masses, the meson decay rates are given by~\cite{Bolton:2024egx}
\begin{equation}
\begin{aligned}
\frac{d\Gamma(P(d_j)\to \tilde{P}(d_j) \psi_k \overline{\psi}_l)}{dq^2}=&\frac{q^2 p_{\tilde{P}}}{24\pi^3} \sqrt{1-\frac{4m_\psi^2}{q^2}} \times\\
\times\Bigg\{& \left[|f_{VV}^{ijkl}|^2\left(1+\frac{2m_\psi^2}{q^2}\right)+|f_{VA}^{ijkl}|^2\left(1-\frac{4m_\psi^2}{q^2}\right)\right]\frac{p_{\tilde{P}}^2}{q^2}f_+^2(q^2) \\
+\frac{3}{8}\frac{(m_{P}^2-m_{\tilde{P}}^2)^2}{m_{P}^2(m_{i}-m_{j})^2}f_0^2(q^2)\Bigg[&|f_{SS}^{ijkl}|^2\left(1-\frac{4m_\psi^2}{q^2}\right)+|f_{SP}^{ijkl}|^2+4|f_{VA}^{ijkl}|^2\frac{m_\psi^2(m_{i}-m_{j})^2}{q^4}\\
+&4\text{Re}\left(f_{VA}^{ijkl}(f_{SP}^{ijkl})^*\right)\frac{m_\psi(m_{j}-m_{i})}{q^2}\Bigg]\Bigg\} \ , \\
\frac{d\Gamma(P(d_j)\to V(d_j) \psi_k \overline{\psi}_l)}{dq^2}=&\frac{q^2 p_{V}}{24\pi^3} \sqrt{1-\frac{4m_\psi^2}{q^2}} \times\\
\times\Bigg\{ 2\frac{p_V^2}{(m_P+m_V)^2} V^2(q^2)&\left[|f_{VV}^{ijkl}|^2\left(1+\frac{2m_\psi^2}{q^2}\right)+|f_{VA}^{ijkl}|^2\left(1-\frac{4m_\psi^2}{q^2}\right)\right] \\
+\left(\frac{(m_P+m_V)^2}{2m_P^2}A_1^2(q^2)+\frac{16m_V^2}{q^2}A_{12}^2(q^2)\right)&\left[|f_{AV}^{ijkl}|^2\left(1+\frac{2m_\psi^2}{q^2}\right)+|f_{AA}^{ijkl}|^2\left(1-\frac{4m_\psi^2}{q^2}\right)\right]\\
+\frac{3}{2}\frac{p_V^2}{(m_{i}+m_{j})^2}A_0^2(q^2)\Bigg[&|f_{PS}^{ijkl}|^2\left(1-\frac{4m_\psi^2}{q^2}\right)+|f_{PP}^{ijkl}|^2+4|f_{AA}^{ijkl}|^2\frac{m_\psi^2(m_{i}+m_{j})^2}{q^4}\\
-&4\text{Re}\left(f_{AA}^{ijkl}(f_{PP}^{ijkl})^*\right)\frac{m_\psi(m_{j}+m_{i})}{q^2}\Bigg]\Bigg\} \ . 
\end{aligned}
\end{equation}
For the decay of $K_L\to\pi^0$, one must substitute
\begin{equation}
\begin{aligned}
f_{SS}^{ijkl} \ & \to \ \sqrt{2}\ \text{Re}\left(f_{SS}^{12kl}\right) \,,\quad
f_{VV}^{ijkl} \ & \to \ \sqrt{2}\ \text{Im}\left(f_{VV}^{12kl}\right) \,, \\
f_{SP}^{ijkl} \ & \to \ \sqrt{2}\ \text{Re}\left(f_{SP}^{12kl}\right) \,,\quad
f_{VA}^{ijkl} \ & \to \ \sqrt{2}\ \text{Im}\left(f_{VA}^{12kl}\right) \,,
\end{aligned}
\end{equation}

\section{Form factors}\label{app:FF}

The form factors, found in appendix~\ref{sec:app_DecayRates} above, are defined as
\begin{equation}
\begin{aligned}
\braket{\tilde{P}(d_i)|\overline{d}_i d_j|P(d_j)}&=\frac{m_P^2-m_{\tilde{P}}^2}{m_j-m_i}f_0(q^2) \,,\\
\braket{\tilde{P}(d_i)|\overline{d}_i \gamma_\mu d_j|P(d_j)}&=\left[(p_P+p_{\tilde{P}})_\mu-q_\mu\frac{m_P^2-m_{\tilde{P}}^2}{q^2}\right]f_+(q^2) +q_\mu\frac{m_P^2-m_{\tilde{P}}^2}{q^2} f_0(q^2) \,,\\
\braket{V(d_i)|\overline{d}_i\gamma_5 d_j|P(d_j)}&=-\frac{2im_V (\epsilon^*\cdot q)}{m_j+m_i}A_0(q^2) \,,\\
\braket{V(d_i)|\overline{d}_i\gamma_\mu d_j|P(d_j)}&=\frac{2\varepsilon_{\mu\nu\alpha\beta}\epsilon^{*\nu}p_P^\alpha p_V^\beta}{m_V+m_P}V(q^2) \,,\\
\braket{V(d_i)|\overline{d}_i\gamma_\mu\gamma_5 d_j|P(d_j)}&=i\epsilon^{*\nu}\Bigg[\frac{2m_Vq_\mu q_\nu}{q^2}(A_0(q_2)-A_3(q_2))+A_1(q^2) g_{\mu\nu}(m_P+m_V) \\
&-A_2(q^2)\frac{(p_P+p_V)_\mu q_\nu}{m_P+m_V}\Bigg] \,,
\end{aligned}
\end{equation}
where $\epsilon$ is the vector polarization of the vector meson and $A_0(0)=A_3(0)$. The form factor $A_3(q^2)$ is given by
\be
A_3(q^2) = \frac{m_B+m_V}{2m_V} A_1(q^2) - \frac{m_B-m_V}{2m_V} A_2(q^2) \ ,
\ee
and it is useful to introduce also
\be
A_{12}(q^2) = \frac{(m_B+m_V)^2(m_B^2-m_V^2-q^2)A_1(q^2)-\lambda(m_B^2,q^2,m_V^2)A_2(q^2)}{16m_Bm_V^2(m_B+m_V)} \ .
\ee

\subsection*{$B\to K$}

The form factors $f_0(q^2)$ and $f_+(q^2)$ for the $B\to K$ transition are taken from Ref.~\cite{Gubernari:2023puw} and, up to isospin symmetry breaking effects, which we neglect here and therefore, are identical for the charged and neutral channels.

\subsection*{$B\to K^{*}$}

The form factors $A_0(q^2)$ and $A_1(q^2)$ for the $B\to K^*$ transition are taken from Ref.~\cite{Gubernari:2023puw}, while $A_{12}(q^2)$ is taken from Ref.~\cite{Bharucha:2015bzk}, and, again, are identical for the charged and neutral channels

\subsection*{$B\to \pi$}

The form factors $f_0(q^2)$ and $f_+(q^2)$ for the $B^0\to \pi^-$ transition are taken from Ref.~\cite{Biswas:2022yvh}. The form factors of the other channels are linked to these by isospin symmetry.

\subsection*{$B\to\rho$}

The form factors $A_1(q^2)$ and $A_2(q^2)$ for the $B^+\to \rho^-$ transition are taken from Ref.~\cite{Bause:2021cna}, while $A_{0}(q^2)$ is taken from Ref.~\cite{Biswas:2022yvh}.
The form factors of the other channels are linked to these by isospin symmetry.

\subsection*{$K\to\pi$}

The form factors $f_0(q^2)$ and $f_+(q^2)$ for the $K^0\to \pi^-$ transition are taken from Ref.~\cite{Carrasco:2016kpy}.
The form factors of the other channels are linked to these by isospin symmetry.

\begin{small}

\bibliographystyle{utphys}
\bibliography{biblio.bib}

\end{small}

\end{document}